\documentclass{physeauth}
\usepackage{graphicx}
\usepackage{scrtime}
\usepackage{latexsym}
\usepackage{cite}
\usepackage{epsfig}
\usepackage{fancyhdr}
\usepackage[hang]{subfigure}
\usepackage{supertabular}
\usepackage{wrapfig}

\begin{document}

\begin{frontmatter}

\title {New Anisotropic Behavior of Quantum Hall Resistance in 
(110) GaAs Heterostructures at mK Temperatures and Fractional Filling Factors}

\author[add1] {Frank Fischer},
\author[add1] {Matthew Grayson\thanksref{ref1}},
\author[add2] {Erwin Schuberth},
\author[add1] {Dieter Schuh},
\author[add1] {Max Bichler} and
\author[add1] {Gerhard Abstreiter}

\address[add1]{Walter Schottky Institut, Technische Universit\"at M\"unchen, 85748,\\
Garching, Germany}

\address[add2]{Walther Meissner Institut, Bayerische Akademie der\\
Wissenschaften, 85748 Garching, Germany}

\thanks[ref1]{
Corresponding author.\\ E-mail: mgrayson@alumni.princeton.edu\\
Phone: +49-89-289-12756\\
Fax: +49-89-3206620
}
 
\begin{abstract}
Transport experiments in high mobility (110) GaAs heterostructures have been
performed at very low temperatures ($8~\textnormal{mK}$).
At higher Landau-Levels we observe a transport anisotropy that bears some similarity with what is already seen
at half-odd-integer filling
on (001) oriented substrates. In addition we report the first observation of
transport anisotropies
within the lowest Landau-Level. This remarkable new anisotropy is independent of the current direction and 
depends on the {\it polarity} of the magnetic field.

\end{abstract}

\end{frontmatter}

\section{Introduction}
Recently there has been increasing interest in transport anisotropies of 
the longitudinal resistance at half-odd-integer filling factors from $\nu=9/2$ upwards
\cite{lil99}.
The anistropy is observed in
high-mobility $(001)$ oriented GaAs/AlGaAs modulation-doped heterostructures with the low resistance
direction typically aligned along  
the $\left[110\right]$ direction, however alignment along  the $\left[ 1 \bar 1 0 \right]$ 
direction has also been observed depending on electron density or 
in-plane magnetic fields \cite{Zhu02,pan99}. To explain this effect, new types of ground states
have been proposed within high Landau levels of the quantum Hall effect (QHE) which are based on
striped phases that align parallel to the low-resistivity direction
\cite{Fog02}.
\section{Sample}
To learn more about the influence of crystal orientation on possible
anisotropies, 
we use a different 
substrate orientation, namely (110) GaAs. The 2-dimensional electron gas (2DEG) is made in a MBE-grown GaAs/AlGaAs $\delta$-doped heterostructure with a spacer 
thickness of $800~\mathring \textnormal{A}$ resulting in peak-mobilities up to {$\mu = 4.2\times 10^{6}~\textnormal{cm}^2/\textnormal{Vs}$} at 
densities $n = 2.1 \times 10^{11}~/\textnormal{cm}^2$. 
In comparison with $(001)$ grown structures,
the two in-plane orthogonal crystal directions on $(110)$, namely $\left[ 1 \bar 1 0 \right]$ 
and $\left[ 001 \right]$,
have explicitly different crystallographic symmetry. \\
\indent For optimizing the mobility of our 2DEGs we grew a variety of substrates where we
varied the temperature, measured by a pyrometer, between $T = 450^\circ \textnormal{C}$ and $T = 530^\circ \textnormal{C}$, with 
$T = 480^\circ \textnormal{C}$ giving the best morphologies and mobilities. The As-flux,
measured as beam equivalent pressure, has been varied between $P_{As} =
3.5\times10^{-5}~\textnormal{mbar}$ and  $P_{As} = 6.5\times10^{-5}~\textnormal{mbar}$, leading to the result
that the highest As-flux gives the best morphologies and 
the highest doping efficiencies of $0.7$ relative to the doping efficiencies on $(001)$. These
findings are in agreement with work previously being done on growth of $(110)$
GaAs \cite{tok98}. 
\begin{figure}[!h]
\center
\includegraphics[width=7cm,keepaspectratio]{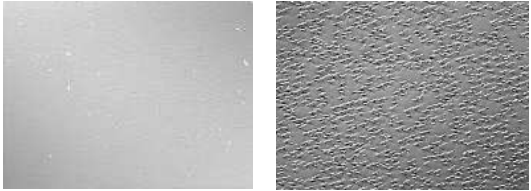}
\caption{Left picture: Surface morphology of a sample grown at $T =480^\circ \textnormal{C}$ and $P_{As}=6\times10^{-5}~\textnormal{mbar}$. 
Right picture: Surface morphology of a sample grown at $T =530^\circ \textnormal{C}$ and $P_{As}=6\times10^{-5}~\textnormal{mbar}$. Scale is $320~\mu \textnormal{m}\times430\mu \textnormal{m}$}
\label{growth}
\end{figure}   
\indent Due to the fragile nature of the anisotropic ground states it is useful to measure at very 
low temperatures, especially in the case of $(110)$ 2DEG's where the mobility is not as high 
as in $(001)$-based systems. The measurements are performed in a dilution refrigerator cryostat 
with a base bath temperature of $T=5~\textnormal{mK}$ and a magnetic field up to $B=7.2~T$.
To have the best possible thermal coupling to the electron system it is crucial
to make the contact resistance as low as possible. For this reason we modified the
usual square-sample van-der-Pauw contacting scheme by
substituting two out of eight contacts with small openings leading to large
contacts with lower resistance and thus, better thermal coupling (Fig. \ref{rxy}
bottom right inset).\\
\section{Experimental results}
\begin{figure}[!h]
\center
\includegraphics[width=7.8cm,keepaspectratio]{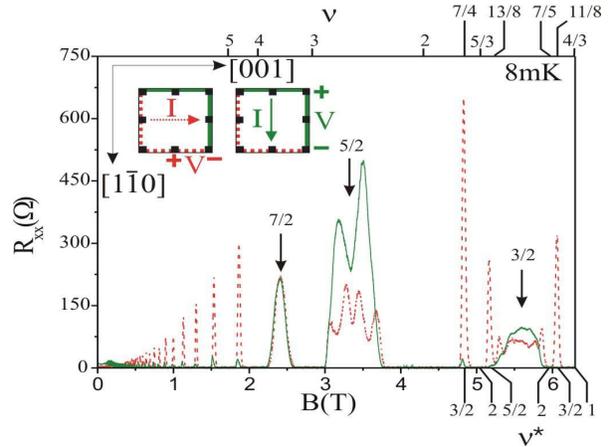}
\caption{Trace of the longitudinal resistance $R_{xx}$ for two perpendicular current directions
indicated in the inset 
at $T=8~\textnormal{mK}$.
As shown in the inset similar colored pairs of $R_{xx}$ voltage contacts show the same type of $R_{xx}$ trace, 
{\it independent of the current contacts used}.
For example all pairs of $R_{xx}$ voltage contacts connected by
the contiguous red dotted line show $R_{xx}$ traces
proportional to the red dotted data trace.
The low temperature
traces develop a strong anisotropy both at $\nu >4$ ($B<2\hspace{1mm}\textnormal{T}$) and strikingly at $\nu <2$
($B>4.5~\textnormal{T}$).}
\label{dashier}
\end{figure}

\begin{figure}[]
\center
\includegraphics[width=7cm,keepaspectratio]{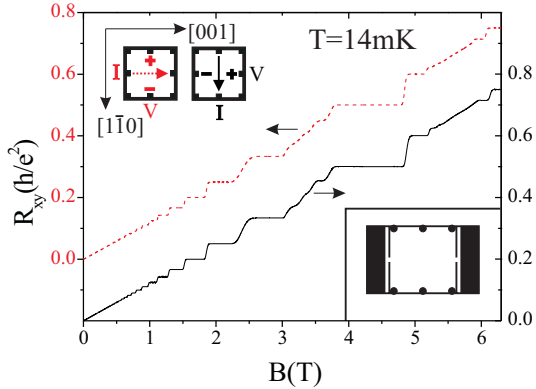}
\caption{Trace of the Hall resistance $R_{xy}$ for two perpendicular current directions
indicated in the inset
at $T=14~\textnormal{mK}$. Inset: van-der-Pauw like sample geometry using 2 large
contacts, on the left and on the right, for improved thermal coupling [6]. }
\label{rxy}
\end{figure}

\begin{figure}[!h]
\center
\includegraphics[width=7cm,keepaspectratio]{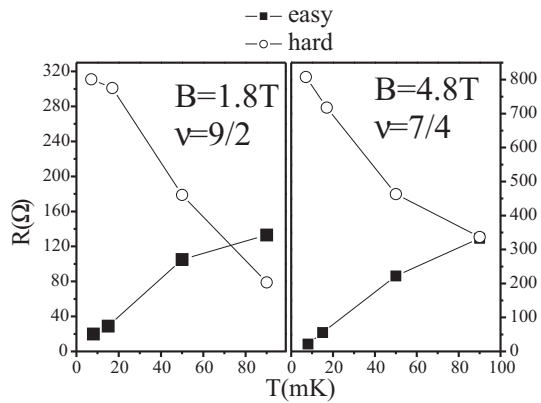}
\caption{Left graph: T-dependence of the $\nu=9/2$ maximum for the contact configuration of Fig. 1. 
Right graph: T-dependence of the maximum at $\nu=7/4$ for the same contacts.}
\label{graph2}
\end{figure}

\indent Fig. \ref{dashier} shows a trace of the longitudinal resistance for the two perpendicular current 
directions, drawn in the inset, at  $T =14~\textnormal{mK}$. 
We examine first the case of {\it exactly} half-filled levels (i.e. the resistance at the center of each peak structure). 
With $\nu < 4$ $(B<2\textnormal{T})$, designated by arrows at exactly $\nu = 7/2, 5/2$ and $3/2$ the resistances in the two directions are 
similar, staying within a factor of 2 of each other, whereas $\nu > 4$ $(B<2\textnormal{T})$ half-filled levels show a highly 
anisotropic behavior differing by a factor of $\sim20$ in resistance, similar to anisotropies in $(001)$ systems. Following 
previous authors we define the low-resistance trace as "easy" and the high-resistance trace as 
the "hard" \cite{pan99}.\\ 
\indent Unlike previous $(001)$ studies, however, a striking anisotropy also develops away from half-filled 
levels in the neighborhood of $\nu = 3/2$ where
fractional quantum Hall states are evident (Fig. \ref{dashier}) in the hard trace. This type of anisotropy has been predicted 
by \cite{lee01}.
It is especially pronounced at filling factors $\nu=7/4$, $13/8$ and $11/8$ which correspond to the
{\it effective half-filled filling factors}
$\nu^*=3/2$, $5/2$ and $3/2$ in the composite fermion picture of the partially filled spin-up Landau level. 
In Fig. \ref{graph2} we see that the temperature dependence of the $\nu = 7/4$ peak in the fractional regime is very similar to the dependence 
of the anisotropy in the $\nu=9/2$ peak in the higher Landau Level regime. We emphasize that all the
resistivity peaks become isotropic again around $100~\textnormal{mK}$. Contrary to the longitudinal resistance
there is no anisotropy to be seen in the Hall resistance (Fig. \ref{rxy}). The traces of the Hall resistance for
 the two prependicular current directions are nearly identical . \\
\indent 
A detailed study indicated that no matter how the current is driven thru the sample, a pair of voltage contacts on the red dotted sample boundary shown 
in the inset
of Fig. \ref{dashier} reveals the "easy" characteristic and a pair of voltage contacts on the green solid sample boundary the "hard".
Unlike previously observed QHE anisotropies, the choice of current
contacts is not relevant.
One of the most remarkable properties of this anisotropy is a dependence on the {\it sign} of the 
magnetic field.
The regions of the sample which show the "hard" resistance characteristic
switch to showing the "easy" characteristic and vice-versa {\it upon reversing the polarity of 
the magnetic field}. An obvious explanation for this behavior is lacking, but it is observed at all voltage contact pairs.

\section{Summary}
We showed that in high-mobility (110) GaAs/AlGaAs heterostructures there is an anisotropy  at very low temperatures.
To our knowledge it is the 
first observation of strong anisotropy of the longitudinal resistance in the fractional quantum Hall regime.
This effect is intriguing in that it cannot be explained by existing theories
for higher Landau levels of electrons, but could possibly be explained
using Landau levels of {\em composite fermions} \cite{lee01}.  This anisotropy,
confirmed in multiple samples from the same wafer, switches character
from "hard" to "easy" and vice-versa upon reversing the polarity of the
magnetic field, while being remarkably insensitive to the direction of
current in the sample.  The switching behavior and the independence of
current direction may suggest an ordered phase with symmetries different
from that of a striped composite Fermion phase proposed in \cite{lee01}.
\ack
This work was supported financially by Deutsche Forschungsgemeinschaft
via Schwerpunktprogramm Quanten-Hall-Systeme \\

\end{document}